\documentclass[pre,aps,showpacs,byrevtex,amsmath,amssymb,onecolumn]{revtex4}
\usepackage{graphicx}
\usepackage{amsmath}
\usepackage{amsfonts}
\usepackage{amssymb}
\begin{document}

\title{Free Energy Formalism for Particle Adsorption}
\author{Pierre Gosselin$^1$ and Herv\'{e} Mohrbach$^2$}
\altaffiliation{Present address: Max-Planck-Institute for Polymer
Research, Theory Group, POBox 3148, D 55021 Mainz, Germany}
\affiliation{$^1$Universit\'{e} Grenoble I, Institut Fourier, UMR\
5582 CNRS-UJF,  UFR de Math\'{e}matiques, BP74, 38402 Saint Martin
d'H\`{e}res, Cedex, France\\
$^2$Laboratoire de Physique Mol\'{e}culaire et des Collisions,
Institut de Physique, Technop\^{o}le 2000, 57078 Metz, France }

%\date{\today}
\begin{abstract}
The equilibrium properties of particles adsorption is investigated
theoretically. The model relies on a free energy formulation which
allows to generalize the Maxwell-Boltzmann description to
solutions for which the bulk volume fraction of potentially
adsorbed particles is very high. As an application we consider the
equilibrium physical adsorption of neutral and charged particles
from solution onto two parallel adsorbing surfaces.
\end{abstract}
\maketitle

\section{Introduction}

The adsorption phenomena, due to the electrochemical interaction
between the particles of a system and a surface, is present in
many experimental set up, such as the adsorption of a perfect gas
on a surface \cite{adamson} or of charged particles in an
electrolyte \cite{israelachvili}. In many physical or chemical
system, a better understanding of the theoretical equilibrium
properties of the adsorbed particles on a surface (see
\cite{barbero1} and references therein) would thus be useful to
interpret the experiments. Several works have been devoted to this
question and various models of particles distribution have been
proposed. Among these, many assumed a Maxwell-Boltzman particle
distribution (see \cite{barbero1}\cite{nazarenko} \cite{kunhau}).
\cite{barbero1}, for instance, study the ionic adsorption on a
surface due to some electrochemical forces in order to determine
the surface density of adsorbed charges versus the thickness of
the sample. This work helps to understand the thickness dependence
of the anisotropic part of the anchoring energy experimentaly
observed \cite{blinov} in a nematic liquid cristal
\cite{nazarenko}\cite{barbero3}\cite{alexe}.

However, a limit can be made about the Maxwell-Boltzman
distribution. Actually, this distribution can only correctly
describe the distribution properties in the dilute regime. But
even in this regime, the density is usually large at the surface
itself, exept when the affinity of the particle for the surface is
weak. To overcome the restriction to the first limit, the dilute
case, we propose to apply a free energy formalism to the study of
the equilibrium properties of neutral and charged particles
adsorption onto two parallel adsorbing surfaces. Another advantage
of the free energy formalism lays in the fact that it leads to the
generalized Poisson-Boltzmann equation introduced in \cite{orland}
which takes into account the finite size of the ions. In that
paper the behavior of electrolytes solutions closed to a charged
surface was studied. In our work, the surface is rather charged by
the adsorption of one of the two charges present in the system.
Within our framework we obtain the electric potential distribution
from the generalized Poisson-Boltzmann equation, the correct
equations for the bulk particle distribution and the density of
the particles on the surface with respect to the thickness $d$ of
the sample. For small thickness, the $d$ dependence of the
electric potential and of the chemical potential are determined
and it is found that the surface density is proportional to
thickness. Whereas in the limit of large $d$ the electric
potential and the surface density are independent of the
thickness. It is nevertheless clear that results obtained with the
phenomenological, coarse-grained free energy formalism to systems
approaching molecular dimension can only be
trusted as far as general trends are concerned.

It is important to note that in the context when several kind of
particles are present in the system, as it would be in an
electrolyte, we find a particle distribution different from the
Fermi-Dirac like distribution introduced in \cite {barbero2}.
Actually, the Fermi-Dirac distribution takes naturally into
account the occupation of the adsorption sites. Yet it misses the
mixing entropy contribution which is present in our formalism. As
a consequence, we show that in the limiting case of a weak
electrolyte, the results of the Poisson-Boltzmann approach are
recovered by our formalism but not by the Fermi-Dirac
distribution.

The paper is organized as follows. In section 2 we introduce the
mean field free energy formalism for neutral particle in a
isotropic fluid limited by two adsorbing surfaces. In the same
section the case of the adsorption competition between two neutral
particles is studied. In section 3 we generalize the free energy
formalism to the study of the ionic adsorption in a isotropic
fluid limited by two adsorbing surfaces, already studied by means
of the Fermi-Dirac distribution in \cite{barbero1} and \cite
{barbero2}.

\section{Neutral particle adsorption}

\bigskip

\subsection{Theory on a lattice}

Consider $N$ neutral particles in a slab of thickness $d$
delimited by two surfaces of area $S$. We divide the slab into
discrete cells of size $a^{3}$ (the size of the particle) and each
cell \ is limited to a single particle occupation. We call $N_{b}$
the number of sites in the bulk and $N_{s}$ the number of surface
adsorption sites. In thermodynamic equilibrium, $n_{b}$ and
$n_{s}$ are the number of particles in the bulk and at the surface
respectively. The volume fraction is then $\phi=n_{b}/N_{b}$ and
the surface density $\phi _{s}=n_{s}/N_{b}$. The conservation of
the total number of particle $N=n_{s}+n_{b}$ is also written:
\begin{equation}
\Phi =2\phi _{s}\frac{a}{d}+\phi (1-\frac{2a}{d})  \label{equi}
\end{equation}
which is valid if the bulk volume fraction is uniform. If this
fraction is not uniform (see section 3), relation (\ref{equi})
becomes:
\begin{equation}
\Phi =2\phi _{s}\frac{a}{d}+\frac{1}{d}\int_{-(d-2a)/2}^{(d-2a)/2}\phi (x)dx
\end{equation}
where $\Phi $ is the total volume fraction.

\subsection{Free energy formalism}

We will use the free energy formalism for the particles
adsorption. It has been first introduced by D. Andelman et al.
(for a review see \cite{ad2}) to describe the kinetic adsorption
of surfactant. This theoretical approach was successfully applied
to the kinetic of non-ionic and ionic surfactant adsorption as
well as to the kinetic of surfactant mixture adsorption. In this
formalism, the two equations describing both the diffusive
transport of surfactant molecules from the bulk solution to the
interface and the kinetic of adsorption at the interface itself
are derived from a single functional. The scope of the present
paper is to apply the free energy formalism to the study of the
equilibrium properties of particles adsorption.

Following \cite{ad2} we write the total free energy as a functional of the
volume fraction in the bulk $\phi (x)$ and the density at the interface $%
\phi _{s}$
\begin{equation}
\frac{F(\phi )}{S}=2f_{s}(\phi _{s})+\int_{-(d-2a)/2}^{(d-2a)/2}f\left[ \phi
\left( x\right) \right] dx
\end{equation}
where the bulk free energy density is written:
\begin{equation}
f(\phi )=\frac{1}{a^{3}}\left\{ kT\left[ (\phi \ln \phi +(1-\phi )\ln \left(
1-\phi \right) \right] -\frac{\widetilde{\beta }}{2}\phi ^{2}-\widetilde{\mu
}\phi \right\}   \label{deltaf}
\end{equation}
and the surface free energy density is equal to :
\begin{equation}
f_{s}\left( \phi _{s}\right) =\frac{1}{a^{2}}\left\{ kT\left[ \phi _{s}\ln
\phi _{s}+\left( 1-\phi _{s}\right) \ln \left( 1-\phi _{s}\right) \right] -%
\widetilde{\alpha }\phi _{s}-\frac{\widetilde{\beta }}{2}\phi _{s}^{2}-%
\widetilde{\mu }\phi _{s}\right\}   \label{deltaf0}
\end{equation}
The parameters $\widetilde{\alpha }$ accounts for the energetic preference
of the particle to absorb on the surface. $\widetilde{\beta }$ is the
lateral interaction between two adjacent charges. Note that the main
difference with the free energy introduced in \cite{ad2} in the context of
surfactant adsorption lies in the presence of the exact entropic term in (%
\ref{deltaf}) rather than on an approximate term. Actually, below as well as
above the critical micellar concentration, the free chain surfactant
solution is always dilute, so that a good approximation for the entropic
term is $S=\phi \ln \phi -\phi $. Another difference is due the effect on
the finite volume: in (\ref{deltaf}) and (\ref{deltaf0}) $\widetilde{\mu }$
is the chemical potential at equilibrium. Its value is not imposed by an
external reservoir but is determined by the conservation equation (\ref{equi}%
). In the infinite volume case considered in \cite{ad2}, the
chemical potential is imposed by an external reservoir localized
at infinity. This last condition imposes the equilibrium bulk
volume fraction.

The variation of $F$ with respect to $\phi (x),$that is
\begin{equation}
\frac{\delta F}{\delta \phi \left( x\right) }=0
\end{equation}
yields the bulk equilibrium volume fraction
\begin{equation}
\phi \left( x\right) =\phi =\frac{1}{1+e^{-\mu -\beta \phi }}
\end{equation}
in which we introduce the dimensional quantities $\beta =\widetilde{\beta }%
/kT$ and $\mu =\widetilde{\mu }/kT$. At the surface the condition
\begin{equation}
\frac{\delta F}{\delta \phi _{s}}=0
\end{equation}
yields the equilibrium adsorption isotherm
\begin{equation}
\phi _{s}=\frac{1}{1+e^{-\mu -\alpha +\beta \phi _{s}}}
\end{equation}
with $\alpha =\widetilde{\alpha }/kT$.

Considering the case $\beta =0$, we find the Fermi-Dirac
distribution that can also be written
\begin{equation}
\phi _{s}=\frac{\phi }{\phi +\left( 1-\phi \right) e^{-\alpha }}
\end{equation}
For $\beta =0$, the number of particles conservation (\ref{equi})
allows to compute analytically the chemical potential :
\begin{equation}
e^{-\mu }=-a+\sqrt{a^{2}+b}  \label{pot1}
\end{equation}
where
\begin{equation}
a=\frac{\left[ \Phi -\left( 1-2a/d\right) \right] +\left( \Phi -2a/d\right)
e^{\alpha }}{2\Phi }  \label{pot2}
\end{equation}
\begin{equation}
b=\frac{\left( 1-\Phi \right) }{\Phi }e^{\alpha }  \label{pot3}
\end{equation}

Note that for $d=2a$ we can check that $e^{-\mu }=e^{\alpha
}\left( 1-\Phi \right) /\Phi $ which leads to the expected result
$\phi _{s}=\Phi $, since all particles are localized on the
surfaces.

\subsubsection{Case $\Phi >>2a/d$}

The condition
\begin{equation}
1>\Phi >>2a/d  \label{condi}
\end{equation}
yields for the chemical potential
\begin{equation}
e^{-\mu }=\frac{\left( 1-\Phi \right) }{\Phi }
\end{equation}
and a bulk volume fraction of :
\begin{equation}
\phi =\Phi .
\end{equation}
Condition (\ref{condi}) implies a negligible variation of the
equilibrium volume fraction after the adsorption process. At the
surfaces,
\begin{equation}
\phi _{s}=\frac{\Phi }{\Phi +\left( 1-\Phi \right) e^{-\alpha }}
\end{equation}
is independent of the size $d$. Even if the surface density of
particles is large, the sample is large enough to ensure that the
bulk volume fraction does not change. This result can also be
obtained from the particle number
conservation equation (\ref{equi}) $\Phi =2\phi _{s}\frac{a}{d}+\phi (1-%
\frac{2a}{d})$ which in the limit $d\rightarrow \infty $ gives $\phi =\Phi $.

\subsubsection{Case $\Phi <<2a/d$}

This condition corresponds to a dilute regime $\Phi <<1$, where $\phi <<1$
so that $\mu \approx \ln \phi $. The surface coverage can then be written
\begin{equation}
\phi _{s}=\frac{\phi }{\phi +e^{-\alpha +\beta \phi _{s}}}
\end{equation}
or from the particle number conservation
\begin{equation}
\phi _{s}=\frac{d\Phi -2a\phi _{s}}{d\Phi -2a\phi _{s}+\left( d-2a\right)
e^{-\left( \alpha +\beta \phi _{s}\right) /T}}  \label{pp}
\end{equation}
The dilute regime for $\beta =0$ is of some interest. For a dilute solution
the chemical potential (\ref{pot1}) is approximatively
\begin{equation}
e^{-\mu }\approx \frac{1+2a/d(e^{\alpha }-1)}{\Phi }
\end{equation}
leading to :
\begin{equation}
\phi \approx \frac{\Phi }{1+2a/d\left( e^{\alpha }-1\right) }  \label{a}
\end{equation}
and, at the surface
\begin{equation}
\phi _{s}\approx \frac{\Phi }{2a/d+e^{-\alpha }\left( 1-2a/d\right) }
\label{b}
\end{equation}
Note that this relation can also be obtained from equation (\ref{pp}).

If $d<<2ae^{\alpha }$, $\phi $ is negligible and (\ref{b}) becomes
\begin{equation}
\phi _{s}\approx \Phi \frac{d}{2a}  \label{phid}
\end{equation}
that is the surface coverage increases linearly with the size of the sample.
Note that in paper \cite{barbero2} the same expression was written
\begin{equation}
\phi _{s}\simeq \frac{N}{2N_{s}}d  \label{dd}
\end{equation}
where $N$ is bulk density of particles in the absence of adsorption and $%
N_{s}$ is the surface density of sites. With the identifications
$\Phi =Na^{3}$ and $N_{s}=1/a^{2}$ the two expressions
(\ref{phid})and (\ref{dd}) coincide.\ But, as the authors of
\cite{barbero2} did not introduce a lattice, expression (\ref{dd})
leads to the unphysical result $\phi
_{s}\rightarrow 0$ in the limit $d\rightarrow 0$, since the correct limit $%
d\rightarrow 2a$ is hidden. Note that this problem will be even more
apparent in the case of the ionic adsorption.

In the opposite limit $d>>2ae^{\alpha }$ from (\ref{a}) we deduce for the
volume fraction in the bulk
\begin{equation}
\phi \approx \frac{\Phi }{1-2a/d}\approx \Phi
\end{equation}
which imposes : $\phi _{s}\frac{2a}{d}<<\Phi $. Actually, from (\ref{b}) we
see
\begin{equation}
\phi _{s}\approx \frac{\Phi }{e^{-\alpha }}
\end{equation}
from which we deduce
\begin{equation}
\phi _{s}\frac{2a}{d}<<\Phi
\end{equation}
In this case the sample is large enough so that the volume
fraction can be considered as constant even when the surface
density is large. The system is then equivalent to an infinite
system coupled to an external reservoir, this last one keeping the
volume fraction constant. We see a crossover between a regime
where the surface coverage increases linearly and another regime
in which the surface coverage is independent of thickness.

\subsection{Equilibrium distribution of two kinds of neutral particles}

In this section, we consider an infinite system composed of two neutral
species which can both adsorb on a flat surface.

We generalize the free energy formulation designed in the preceding section
by writing the bulk contribution of the density free energy in $kT$ units
\begin{eqnarray}
f(\phi _{A},\phi _{B}) &=&\frac{kT}{a^{3}}\left\{ \phi _{A}\ln \phi
_{A}+\phi _{B}\ln \phi _{B}+(1-\phi _{A}-\phi _{B})\ln \left( 1-\phi
_{A}-\phi _{B}\right) \right.  \\
&&\left. -\frac{\beta _{A}}{2}\phi _{A}^{2}-\frac{\beta _{B}}{2}\phi
_{B}^{2}-\varepsilon \phi _{A}\phi _{B}-\mu _{A}\phi _{A}-\mu _{B}\phi
_{B}\right\}
\end{eqnarray}
where $\varepsilon $ is an interaction between the two species. At the
surface, we have :
\begin{eqnarray}
f_{s}(\phi _{s,A},\phi _{s,B}) &=&\frac{kT}{a^{3}}\left\{ \phi _{s,A}\ln
\phi _{s,A}+\phi _{s,B}\ln \phi _{s,B}+(1-\phi _{s,A}-\phi _{s,B})\ln
(1-\right.  \\
&&\phi _{s,A}-\phi _{s,B})-\alpha _{A}\phi _{s,A}-\alpha _{B}\phi _{s,B}-%
\frac{\beta _{A}}{2}\phi _{s,A}^{2}-\frac{\beta _{B}}{2}\phi _{s,B}^{2} \\
&&\left. -\varepsilon \phi _{s,A}\phi _{s,B}-\mu _{A}\phi _{s,A}-\mu
_{B}\phi _{s,B}\right\}
\end{eqnarray}
Note the presence of the mixing entropic term $(1-\phi _{A}-\phi
_{B})\ln \left( 1-\phi _{A}-\phi _{B}\right) $ in these two
expressions. This terms is very important since it avoids that two
particles of different kind sit at the same place in the lattice.
Its absence would lead to the FD distribution.

Minimizing the free energy, we obtain in the bulk:
\begin{equation}
\phi _{A}=\frac{1-\phi _{B}}{1+e^{-(\mu _{A}+\beta _{A}\phi _{A}+\varepsilon
\phi _{B})}}
\end{equation}
and
\begin{equation}
\phi _{B}=\frac{1-\phi _{A}}{1+e^{-(\mu _{B}+\beta _{B}\phi _{B}+\varepsilon
\phi _{A})}}
\end{equation}
whereas at the surface we have:
\begin{equation}
\phi _{s,A}=\frac{1-\phi _{s,B}}{1+e^{-(\mu _{A}+\alpha _{A}+\beta _{A}\phi
_{s,A}+\varepsilon \phi _{s,B})}}
\end{equation}
and
\begin{equation}
\phi _{s,B}=\frac{1-\phi _{s,A}}{1+e^{-(\mu _{B}+\alpha _{B}+\beta _{B}\phi
_{s,B}+\varepsilon \phi _{s,A})}}
\end{equation}
We thus see that the distributions of the two species are not independent of
each other, due to the mixing entropy.

Suppose now that $\alpha _{A}>>\alpha _{B}$. We find
\begin{equation}
\phi _{s,B}\approx \frac{e^{\mu _{B}+\alpha _{B}}}{e^{\mu _{A}+\alpha _{A}}}%
<<1
\end{equation}
and
\begin{equation}
\phi _{s,A}=\frac{1}{1+e^{-\mu _{A}-\alpha _{A}}}
\end{equation}
showing that only one specie adsorbs, the other staying in the bulk. One can
check that $\phi _{s,A}+\phi _{s,B}$ is always smaller than one.

Now, let us compare our result with the Fermi-Dirac distribution. In such a
context, the distribution for the two species are:
\begin{equation}
\phi _{A,B}=\frac{1}{1+e^{-\mu _{A,B}}}
\end{equation}
for the bulk and
\begin{equation}
\phi _{s,A,B}=\frac{1}{1+e^{-\mu _{A,B}-\alpha _{A,B}}}
\end{equation}
for the surface. The two distributions are now completely independent. In
particular for the $\alpha _{A}>>\alpha _{B}$ the sum $\phi _{s,A}+\phi
_{s,B}$ is not guarantee to be smaller than $1$. This example shows the
importance of taking the mixing entropy into account when more than one
specie are present.

\bigskip

\section{Ionic adsorption}

The power of the free energy formalism can also be applied to the ion
distribution in an isotropic fluid limited by two adsorbing surfaces. As
explained in \cite{barbero1} and \cite{barbero2}, this system has already
been considered by several authors. Actually, the ionic adsorption has been
invoked to explain the thickness dependence of the anisotropic part of the
anchoring energy of the interface between a substrate and a nematic liquid
crystal.

Consider a slab of thickness $d$ with two identical adsorbing flat surfaces
that adsorb only positive ions. Obviously the liquid is globally neutral.
However, due to the selective ionic adsorption there is a distribution of
charges yielding a non uniform electric potential $V\left( x\right) $ across
the sample. Since the surfaces are identical i.e. the affinity of the
positive ions for the surfaces are identical, the potential is symmetric $%
V\left( x\right) =V\left( -x\right) $ and $E=-dV/dx$ is vanishing at the
middle of the sample.

The total free energy for a symmetric electrolyte in the mean
field approximation has already been introduced in \cite{orland}
in the context of the adsorption of large ions from a solution of
infinite size to a charged surface. In our case it is rather the
adsorption phenomena which charges the surfaces. Within the
mean-field approximation, the total free energy in the bulk
$f=u-Ts$ can be written in terms of the local electrostatic
potential in $kT$ unit $\psi (x)=eV\left( x\right) /kT$ and the
ions volume fraction $\phi ^{\pm }\left( x\right) $. The
electrostatic energy contribution is
\begin{equation}
u=\frac{kT}{a^{3}}\int dx\left[ -L_{B}^{2}\left| \frac{\partial \psi }{%
\partial x}\right| ^{2}+\phi ^{+}\psi -\phi ^{-}\psi -\mu _{+}\phi ^{+}-\mu
_{-}\phi ^{-}\right]   \label{u}
\end{equation}
where $\varepsilon $ is the dielectric constant of the solution,
$\mu _{\pm }$
are the equilibrium chemical potential of the two ions and $L_{B}=\sqrt{%
\frac{\varepsilon kTa^{3}}{2e^{2}}}$ is the intrinsic length of the problem.
Note that we use the same system of unit as \cite{barbero2} which is
different from the one of \cite{orland} where $L_{B}=\sqrt{\frac{\varepsilon
kTa^{3}}{8\pi e^{2}}}$. The first term in the left hand side of (\ref{u}) is
the self energy of the electric field, the next two terms are the
electrostatic energy of the ions. For the sake of simplicity we do not
introduce additional steric interaction. The entropic contribution is
\begin{equation}
Ts=-\frac{kT}{a^{3}}\int dx\left[ \phi _{+}\ln \phi _{+}+\phi _{-}\ln \phi
_{-}+(1-\phi _{+}-\phi _{-})\ln \left( 1-\phi _{+}-\phi _{-}\right) \right]
\end{equation}
The first two terms represent the tanslational entropy of the ions and the
last term the entropy of mixing i.e. the entropy of the solvant molecules.

At the interface itself, the total free energy is obtained by adding an
electrostatic contribution $\phi _{s}^{+}\psi _{s}$ to equation (\ref
{deltaf0})
\begin{eqnarray}
f_{s}\left( \phi _{s}^{+},\psi _{s}\right)  &=&\frac{kT}{a^{2}}\left\{ \left[
\phi _{s}^{+}\ln \phi _{s}^{+}+\left( 1-\phi _{s}^{+}\right) \ln \left(
1-\phi _{s}^{+}\right) \right] -\alpha \phi _{s}^{+}+\phi _{s}^{+}\psi
_{s}\right.   \notag \\
&&\left. -\mu ^{+}\phi _{s}^{+}\right\}
\end{eqnarray}
\ The total free energy of the system is then:
\begin{equation}
\frac{F(\phi ^{\pm })}{S}=2f_{s}(\phi _{s}^{+},\psi
_{s})+\int_{-(d-2a)/2}^{(d-2a)/2}f\left[ \phi ^{\pm }\left( x\right) ,\psi
\left( x\right) \right] dx
\end{equation}
The ions number conservation imposes the equality between the two chemical
potentials $\mu _{+}=\mu _{-}=\mu $. The variation of the bulk free energy
with respect to $\phi ^{\pm }$ yields the volume fraction of the ions in the
bulk:
\begin{equation}
\phi ^{+}=\frac{e^{-\psi +\mu }}{h(\psi ,\mu )}
\end{equation}
and
\begin{equation}
\phi ^{-}=\frac{e^{\psi +\mu }}{h(\psi ,\mu )}
\end{equation}
where
\begin{equation}
h(\psi ,\mu )=1+2e^{\mu }\cosh \psi
\end{equation}
Note that the distribution in the bulk is very different from the
FD\ distribution which reads
\begin{equation}
\phi _{FD}^{\pm }=\frac{1}{1+e^{-\mu \pm \psi }}  \label{FD}
\end{equation}
The variation of the bulk free energy with respect to $\psi $ yields the
modified Poisson Boltzman equation introduced in \cite{orland} :
\begin{equation}
\nabla ^{2}\psi =\frac{e^{\mu }}{L_{B}^{2}}\frac{\sinh \psi }{h(\psi ,\mu )}
\end{equation}
This equation is similar to the expression given in \cite{barbero2}. The
difference lies in the formula for $h(\psi ,\mu )$ which is in \cite
{barbero2} : $h(\psi ,\mu )=1+2e^{\mu }\cosh \psi +e^{-2\mu }.$

The variation of the total free energy with respect to $\psi \left(
x=d/2\right) \equiv \psi _{s}$ yields the requirement of the overall charge
neutrality:
\begin{equation}
\left. \frac{\partial \psi }{\partial x}\right| _{x=d/2}=\frac{-1}{L_{s}}%
\phi _{s}
\end{equation}
where $L_{s}=a^{2}\varepsilon kT/e^{2}$ is a caracteristic length of the
surface introduced in \cite{barbero2}. Minimizing the surface free energy $%
\phi _{s}$ yields
\begin{equation}
\phi _{s}=\frac{1}{1+e^{\psi -\alpha -\mu }}  \label{phis}
\end{equation}
which is a Fermi-Dirac distribution.

Note that a FD distribution is obtained for $\phi ^{-}$ when the
electrostatic potential is very high $\psi >>1$ since in this case:
\begin{equation}
\phi ^{-}\rightarrow \frac{1}{1+e^{-\psi -\mu }}
\end{equation}
whereas
\begin{equation}
\phi ^{+}\rightarrow e^{-2\psi }\phi ^{-}\neq \frac{1}{1+e^{\psi -\mu }}
\end{equation}

It is instructive to analyze the behaviour of the system for semi infinite
and very thin sample and\ then to compare the prediction of our model to the
results obtained with the FD distribution.

\subsection{Infinite volume limite}

\bigskip In the infinite volume limit $d\rightarrow \infty $, we have :
\begin{equation}
e^{\mu }=\frac{\phi _{0}}{2(1-\phi _{0})}
\end{equation}
whereas
\begin{equation}
e_{FD}^{\mu }=\frac{\phi _{0}}{2-\phi _{0}}
\end{equation}
due to the lack of the mixing entropy. The generalized PB equation can be
solved numerically to find the electric field and the ions distributions
across the sample. This has been done in \cite{orland} where interesting
curves can be found.

It is only in the dilute case that the two chemical potential coincide and
are equal to the Boltzman one :
\begin{equation}
e_{\text{Boltzman}}^{\mu }=\frac{\phi _{0}}{2}
\end{equation}
In this case, we obtain the surface potential
\begin{equation}
\psi _{s}=\frac{2\alpha }{3}+\frac{2}{3}\ln \left( \frac{L_{B}}{L_{s}}\sqrt{%
\frac{\phi _{0}}{2}}\right)
\end{equation}
and the surface coverage :
\begin{equation}
\phi _{s}=\frac{1}{1+e^{\alpha /3}\left( \frac{L_{B}}{L_{s}}\frac{2}{\phi
_{0}}\right) ^{2/3}}
\end{equation}
in agreement with \cite{barbero2}.

\subsubsection{Small volume limit}

In the finite volume case we have two conservation laws
\begin{equation}
\Phi =\frac{2a}{d}\phi _{s}+\frac{1}{d}\int_{-\left( d-2a\right) /2}^{\left(
d-2a\right) /2}\frac{e^{\mu -\psi }}{1+2e^{\mu }ch(\psi )}  \label{c1}
\end{equation}

\begin{equation}
\Phi =\frac{1}{d}\int_{-\left( d-2a\right) /2}^{\left( d-2a\right) /2}\frac{%
e^{\mu +\psi }}{1+2e^{\mu }ch(\psi )}  \label{c2}
\end{equation}
Let us consider the small volume limit $d\rightarrow 3a$.
Physically, we cannot consider a smaller bound since the negative
charges are not adsorbed\ \ In particular, it is not possible to
take the limit $d\rightarrow 0$.

The relation $\left( \ref{c2}\right) $ becomes (assuming $e^{\psi _{0}}$ $%
\gg 1$, which will be justified later):
\begin{equation}
\Phi \approx \frac{d-2a}{d}\frac{e^{\mu +\psi _{0}}}{1+e^{\mu +\psi _{0}}}
\end{equation}
which leads to :
\begin{equation}
e^{\mu }=\frac{\Phi }{1-\frac{2a}{d}-\Phi }e^{-\psi _{0}}
\end{equation}
Considering the approximation :

\begin{equation}
\psi _{s}\approx \psi _{0}-\frac{a}{L_{s}}\phi _{s}
\end{equation}
and the fact that $\frac{a}{L_{s}}$ is very small, we can assume that :
\begin{equation}
\psi _{s}\approx \psi _{0}
\end{equation}
in $\left( \ref{phis}\right) $, so that :
\begin{equation}
\phi _{s}\approx \frac{\Phi }{\Phi +\left( 1-\frac{2a}{d}-\Phi \right)
e^{-\alpha +2\psi _{0}}}\text{.}
\end{equation}
Plugging this result in $\left( \ref{c1}\right) $ and using the fact that $%
e^{\alpha }\gg 1$ yields the electric potential :
\begin{equation}
e^{\psi _{0}}\approx e^{\alpha /2}\sqrt{\frac{\frac{2a}{d}-\Phi }{1-\frac{2a%
}{d}-\Phi }}  \label{psi0}
\end{equation}
confirming our assumption $e^{\psi _{0}}\gg 1$.

For the chemical potential we readily obtain:

\begin{equation}
e^{\mu }\approx \frac{\Phi }{1-\frac{2a}{d}-\Phi }\sqrt{\frac{1-\frac{2a}{d}%
-\Phi }{\frac{2a}{d}-\Phi }}e^{-\alpha /2}\text{.}  \label{mu}
\end{equation}
and for the surface coverage we find :
\begin{equation}
\phi _{s}\approx \frac{d\Phi }{2a}\text{.}
\end{equation}
These results are very different from the one obtained by Barbero
et al. with the Fermi Dirac distribution which are :
\begin{equation}
e^{\psi _{0}}\sim \frac{e^{\alpha /2}}{d}
\end{equation}
\begin{equation}
e^{\mu }\sim \sqrt{\Phi }
\end{equation}
These two last relations do not lead to the result obtained with the
Maxwell-Boltzman distribution used in the dilute regime. In other words, the
Fermi- Dirac distribution in the limit of a small concentration does not
lead to the correct Maxwell- Boltzman result.

On the contratry, our equations $\left( \ref{psi0}\right) $ and $\left( \ref
{mu}\right) $, for $\Phi \ll 1$, yields :
\begin{equation}
e^{\psi _{0}}\approx e^{\alpha /2}\sqrt{\frac{2a}{d-2a}}
\end{equation}
and
\begin{equation}
e^{\mu }\approx \sqrt{\frac{d^{2}}{2a\left( d-2a\right) }}\Phi e^{-\alpha /2}
\end{equation}
that are the results obtained with a Maxwell-Boltzmann
distribution.

An estimation of the parameter was given in \cite{barbero2} for a
typical nematic liquid crystal ($\epsilon\approx 6$ for an organic
liquid) limited by two glasses. The adsorption energy was
evaluated $\alpha\approx 6$ and for a typical molecule of radius
$R\approx 40 $A one has $L_B\approx 30 $A. The surface density was
found to be $d$ dependent for thickness smaller than $300$ A.

\section{conclusion}

In this paper, we have proposed a free energy formalism to
describe the phenomenon of surface adsorption of neutral and
charged particles as well. This free energy formalism has led to
the equilibrium particles distribution for the case of physical
adsorption of neutral and charged particles from solution onto two
parallel adsorbing surfaces. In particular, we have found the
correct equations for the electric potential and the equilibrium
charge distribution with respect to the thickness of the
electrolyte sample in case of high bulk concentration and we
recover the results obtained with the Maxwell-Boltzmann
distribution in the limit of small concentration.

We are aware that our model relies on some strong assumptions, in
particular the adsorbed particles are confined to a monomolecular
layer whereas multilayer adsorption is frequently observed.
Moreover, we have assumed that the surface is homogenous which is
obviously not the case in general.

Nevertheless,one of the advantages of the free energy formalism is that it
relies on a minimization principle, avoiding in this way the introduction of
ad-hoc distributions, and allowing straightforwardly the description of
multi particles adsorption. Moreover it can be extended to take into account
the particles interactions in the bulk and at the surface itself, that are
usually not considered. This generalization needs further investigation.

\end{document}